# Theoretical analysis on the transient ignition of premixed expanding flame in a quiescent mixture

Dehai Yu and Zheng Chen[1]

BIC-ESAT, SKLTCS, CAPT, College of Engineering, Peking University, Beijing 100871, China

**Abstract**

The ignition of a self-sustained premixed expanding flame is an important process in various combustion facilities and constitutes a crucial problem in fundamental combustion research. In this work, a transient formulation on the forced ignition of premixed expanding spherical flame in a quiescent mixture is proposed. The present theory considers the transient evolution of the temperature and fuel mass fraction distributions, and it can determine the critical heating power and minimum ignition energy for successful ignition. The flame initiation process subject to central heating can be approximately decomposed into four stages, i.e., the fast establishment of the ignition kernel, the ignition-energy-supported flame kernel propagation, unsteady transition of the flame kernel, and quasi-steady spherical flame propagation. Comparison between present transient theory and previous quasi-steady theory indicates that the unsteady effects directly lead to the appearance of flame kernel establishing stage and considerably affect the subsequent flame kernel development by lowering the flame propagation speed. Time scale analysis is conducted, and it is found that the transient formulation completely degenerates to the quasi-steady theory either in the neighborhood of stationary flame ball or as approaching the planar flame. Previous quasi-steady theory shows that the critical heating power for successful ignition is proportional to the cube of the critical flame radius, i.e., $Q_{cr} \sim R_{cr}^3$. However, the critical heating power predicted by the present transient formulation deviates this scaling law, which can be attributed to the unsteady thermal conduction from heating center to the flame front. Furthermore, in the transient formulation the central heating with finite duration is considered and the minimum ignition energy is obtained. For the first time, the memory effect that persistently drives flame propagation subsequent to central heating switching-off is examined. It is found that as the heating power grows, the memory effect becomes increasingly

---

[1] Corresponding author, Email address: cz@pku.edu.cn, Tel: 86-10-62766232

important and it can greatly reduce the minimum ignition energy.

**Key words**: Ignition, spherical flame, unsteady effect, critical radius, minimum ignition energy

# 1. Introduction

Flame initiation or forced ignition in a flammable mixture refers to the generation of a self-sustained propagating flame front from an ignition kernel. Flame initiation plays an important role in fundamental combustion research. Besides, understanding ignition is important for controlling ignition in advanced engines and preventing fire or explosion. In general, forced ignition is triggered by the deposition of certain amount of thermal energy, such as electric spark or a hot solid body, which rises the local temperature and incudes intensive chemical reaction and thermal runaway (Joulin 1985, Ronney 1990, He 2000, Chen & Ju 2007).

Successful ignition is achieved only when the heat generation from chemical reaction overcomes heat loss to the surrounding environment. Adopting large activation energy asymptotics, Vázquez-Espí and Liñán (2001, 2002) analyzed the ignition characteristics of a gaseous mixture subject to a point energy source. They identified two ignition regimes through comparing the relevant time scales including the homogeneous ignition time ($t_{ch}$), the characteristic time for acoustic wave propagation ($t_a$), and the characteristic time for heat conduction ($t_c$). The ratio $t_a/t_c$ is equivalent to the Knudsen number and it is typically quite small i.e. $t_a \ll t_c$. The first regime is for fast ignition energy deposition with $t_{ch} \cong t_a \ll t_c$. In this regime, the heat loss due to thermal expansion balances the heat release from chemical reaction. The second regime is for moderate ignition energy deposition with the corresponding reaction rate being comparable to the heat conduction rate, i.e., $t_a \ll t_{ch} \cong t_c$. As the pressure wave passes across the hot spot, the local chemical reaction proceeds slightly. This regime corresponds to the diffusive ignition occurring under near isobaric condition. In this study, we consider the ignition process in the second regime.

During the ignition process, the reactant consumption becomes relevant and thus the ignition kernel development is affected by the diffusive properties of the deficient reactant. Previous studies have vastly investigated the diffusion-controlled premixed stationary spherical flame, which is also known as flame ball and closely related to ignition (Ya et al. 1980, Ronney 1989). Based on the thermal-diffusion model, Deshaies and Joulin (1984) conducted linear stability analysis and found that the adiabatic flame ball is absolutely unstable. This indicates that a negative perturbation of flame radius results in inward collapse and subsequent flame extinction, while a positive displacement perturbation leads to outward propagation. Therefore, the flame ball radius is popularly considered as the critical radius for successful ignition, beyond which the flame kernel can spontaneously evolve into a self-sustained flame (Chen & Ju 2007, Kelley et al. 2009). However, in premixtures with high Lewis numbers ($Le > 1$), the critical radius for successful ignition is in fact much smaller than the flame ball radius (He 2000, Chen et al. 2011). Consequently, the minimum ignition energy (MIE) could be greatly over-predicted based on the flame ball radius.

Practically ignition is usually triggered by the energy deposition, which can be approximately modeled as continuous central heating (Deshaies & Joulin 1984, Jackson et al. 1989). When the heating power is sufficiently low, ignition fails and the self-sustained expanding flame cannot be achieved (Deshaies & Joulin 1984, Chen & Ju 2007). Successful ignition is achieved only when the heating power is high enough to induce a continuous transition from flame kernel to self-sustained expanding flame. Once the flame kernel evolves in a self-sustained manner, the central heating becomes irrelevant and could be switched off after an appropriate duration of time. This yields a finite amount of energy deposition, and thereby we can determine the MIE (Chen et al. 2011, Fernández-Tarrazo et al. 2016). In the course of external heating, the characteristics of the flame front, e.g., flame temperature, flame propagation speed, and flame curvature/stretch, undergo substantial change. This implies the necessity in taking account of the unsteady effects on the ignition process (Kurdyumov et al. 2004, Chen et al. 2011). According to He (2000), the duration for the flame to reach the critical radius $t_h$ can be evaluated by a nonlinear velocity-curvature relation derived based on quasi-steady assumption. The product of $t_h$ with heating power $Q_s$ gives the MIE. Employing the thermal-diffusion model and associating with numerical simulation, Chen et al. (2011) suggested that the MIE tends to be linearly proportional to the cube of the critical flame radius. However, the quasi-steady assumption implies that the system is fully developed, and correspondingly the temperature and mass fraction profiles across the reaction front are given by their final state after the long-term evolution. This quasi-steady assumption might not be suitable for the initial ignition kernel development.

Lack of characteristic timescale, the quasi-steady theory cannot rigorously interpret the dynamic behavior of the flame kernel development subsequent to the switching off the heating source. For instance, the removal of heating source would not cause immediate flame quench; instead the flame could propagate for a finite distance due to the memory effect (Joulin 1985, He 2000). Employing large activation energy asymptotics, Joulin (1985) analyzed the point-source-induced flame initiation in lean light reactant with *Le* < 1. A nonlinear evolution equation describing the time-dependent evolution of the flame kernel and determining the MIE was derived through matching the asymptotic solutions for the quasi-steady zone close to the flame ball radius and for the unsteady far-field (Joulin 1985, Buckmaster & Joulin 1989). However, the neglect of convection-related terms in the asymptotic analysis for both near-field quasi-steady zone and far-field unsteady zone might restrict the validity domain of the evolution equation, and thus the evaluation of the MIE. Besides, for mixtures with large Lewis number, the flame ball size is considerably larger than the critical radius and thus tends to be irrelevant to flame initiation (Chen et al. 2011). Employing asymptotic analysis, Clavin (2017) described the dynamic quenching of spherical flame expanding at large radius beyond flammability limits of planar flames, which has been observed in micro-gravity experiments (Ronney 1989, Ronney 1990). However, the unsteady effect was not considered by Clavin (2017).

Sensible evaluation of MIE requires analyzing the propagation mechanism of ignited flame kernel. The unsteady effect characterizing the time change of temperature and mass fraction across the flame front is expected to have direct impact upon the flame propagation dynamics. However, the unsteady effect has not been clarified in previous theoretical studies. This work aims to develop a fully transient formulation describing the flame initiation process. It generalizes the quasi-steady theory by rigorously taking unsteady effects into account and is valid over the entire spatial domain for flame initiation. The transient formulation can be used to assess the unsteady effect on ignition kernel propagation and MIE

The paper is organized as follows. In section 2, the transient formulation is proposed and solved analytically. The analytical solutions for the time-dependent temperature and reactant mass fraction distributions on each side of the flame front are obtained. The solutions describing the temporal evolution of flame temperature and flame propagation speed are obtained from matching conditions. In section 3, a thorough comparison between the transient formulation and the quasi-steady theory is presented with emphasis on the dynamic behavior of flame front propagation, the evaluation of critical heating power and minimum ignition energy, and the assessment of the memory effect. The concluding remarks are given in section 4.

## 2. Formulation

### 2.1 Governing Equations

Energy deposition into a combustible mixture increases the local temperature and generates an ignition kernel subsequently. For simplicity, we consider the development of the ignition kernel in a quiescent mixture under microgravity condition. Providing that the stoichiometric ratio of the mixture is far away from the flammability limit, the impacts of radiation heat loss upon the propagation of the flame kernel appears to be quantitative instead of qualitative (Chen & Ju 2008, Chen 2017). Therefore, the effect of radiative loss is not considered here, and it can be explored in future studies. We use the classical thermal-diffusive model, in which the density $\tilde{\rho}$, heat capacity $\tilde{C}_p$, thermal conductivity $\tilde{\lambda}$, mass diffusion coefficient of the deficient reactant $\tilde{D}$, and heat of reaction $\tilde{q}$ are assumed to be constant. In the thermal-diffusive model, the thermal expansion or convective effect is not considered. According to Champion et al. (1988), thermal expansion only quantitatively affects the MIE and the key features of ignition are covered by using the thermal-diffusive model. Besides, we also assume an overall one-step exothermic reaction. These assumptions have been widely adopted in previous theoretical studies (He 2000, Chen & Ju 2007). The governing equations for temperature, $\tilde{T}$, and mass fraction of the deficient reactant, $\tilde{Y}$, are

$$\tilde{\rho}\tilde{C}_p \frac{\partial \tilde{T}}{\partial \tilde{t}} = \frac{1}{\tilde{r}^2}\frac{\partial}{\partial \tilde{r}}\left(\tilde{r}^2 \tilde{\lambda}\frac{\partial \tilde{T}}{\partial \tilde{r}}\right) + \tilde{q}\tilde{\omega} \qquad (1)$$

$$\tilde{\rho}\frac{\partial \tilde{Y}}{\partial \tilde{t}} = \frac{1}{\tilde{r}^2}\frac{\partial}{\partial \tilde{r}}\left(\tilde{r}^2 \tilde{\rho}\tilde{D}\frac{\partial \tilde{Y}}{\partial \tilde{r}}\right) - \tilde{\omega} \qquad (2)$$

where $\tilde{t}$ and $\tilde{r}$ are the time and radial coordinate, respectively. The reaction rate follows the Arrhenius law as

$$\tilde{\omega} = \tilde{\rho}\tilde{A}\tilde{Y}\exp\left(-\frac{\tilde{E}_a}{\tilde{R}^0 \tilde{T}}\right) \qquad (3)$$

where $\tilde{A}$ is the pre-factor, $\tilde{E}_a$ the activation energy, and $\tilde{R}^0$ the universal gas constant.

The laminar flame speed $\tilde{S}_L^0$, flame thickness $\tilde{\delta}_L^0 = \tilde{\lambda}/(\tilde{\rho}\tilde{c}_p\tilde{S}_L^0)$, and characteristic flame time $\tilde{t}_L^0 = \tilde{\delta}_L^0/\tilde{S}_L^0$ for the adiabatic planar flame are used as the reference speed, length and time, respectively. The non-dimensional quantities are defined as

$$u = \tilde{u}/\tilde{S}_L^0, \qquad r = \tilde{r}/\tilde{\delta}_L^0, \qquad t = \tilde{t}/\tilde{t}_L^0 \qquad (4)$$

In addition, the normalized temperature and mass fraction are defined by

$$T = \frac{\tilde{T} - \tilde{T}_\infty}{\tilde{T}_{ad} - \tilde{T}_\infty}, \qquad Y = \frac{\tilde{Y}}{\tilde{Y}_\infty} \tag{5}$$

where $\tilde{T}_\infty$ and $\tilde{Y}_\infty$ are respectively the temperature and mass fraction of the deficient reactant of the unburned mixture. The adiabatic flame temperature is $\tilde{T}_{ad} = \tilde{T}_\infty + \tilde{Y}_\infty \tilde{q}/\tilde{c}_p$.

The non-dimensional form for the governing equations (1) and (2) is

$$\frac{\partial T}{\partial t} = \frac{1}{r^2}\frac{\partial}{\partial r}\left(r^2 \frac{\partial T}{\partial r}\right) + \omega \tag{6}$$

$$\frac{\partial Y}{\partial t} = \frac{1}{Le}\frac{1}{r^2}\frac{\partial}{\partial r}\left(r^2 \frac{\partial Y}{\partial r}\right) - \omega \tag{7}$$

where $Le = \tilde{\lambda}/(\tilde{\rho}\tilde{c}_p\tilde{D})$ is the Lewis number. The non-dimensional chemical reaction rate is $\omega = \tilde{\delta}_L^0 \tilde{\omega}/(\tilde{\rho}\tilde{S}_L^0 \tilde{Y}_\infty)$. The parameters with and without carrot symbol denote the dimensional and non-dimensional variables, respectively.

In the limit of large activation energy, the reaction zone appears to be infinitely thin and the reaction rate can be modeled by a delta function located at the reaction zone (Law 2006, Veeraragavan & Cadou 2011, Wu & Chen 2012), i.e.,

$$\omega = \left[\epsilon_T + (1-\epsilon_T)T_f\right]^2 \exp\left[\frac{Z}{2}\frac{T_f - 1}{\epsilon_T + (1-\epsilon_T)T_f}\right]\delta(r - R) \tag{8}$$

where $T_f$ is the normalized flame temperature, $R$ the flame front position (or flame radius), $Z = \tilde{E}_a(1-\epsilon_T)/\tilde{R}^0 \tilde{T}_{ad}$ the Zel'dovich number, and $\epsilon_T = \tilde{T}_\infty/\tilde{T}_{ad}$ the expansion ratio.

The flame front separates the unburnt and burnt regions. In these two regions, the reaction term does not appear in the governing equations. Therefore, the governing equations can be written in the burnt and unburnt regions as

*Burnt region*

$$\frac{\partial T_b}{\partial t} = \frac{1}{r^2}\frac{\partial}{\partial r}\left(r^2 \frac{\partial T_b}{\partial r}\right) \tag{9}$$

$$\frac{\partial Y_b}{\partial t} = \frac{1}{Le}\frac{1}{r^2}\frac{\partial}{\partial r}\left(r^2 \frac{\partial Y_b}{\partial r}\right) \tag{10}$$

*Unburnt region*

$$\frac{\partial T_u}{\partial t} = \frac{1}{r^2}\frac{\partial}{\partial r}\left(r^2 \frac{\partial T_u}{\partial r}\right) \tag{11}$$

$$\frac{\partial Y_u}{\partial t} = \frac{1}{Le}\frac{1}{r^2}\frac{\partial}{\partial r}\left(r^2 \frac{\partial Y_u}{\partial r}\right) \tag{12}$$

where the subscripts $u$ and $b$ represent states in the unburnt and burnt regimes, respectively.

The initial and boundary conditions can be written as

$$t = 0: \quad T_b = T_f \ \& \ Y_b = 0 \quad \text{for } r \leq R(t), \qquad T_u = 0 \ \& \ Y_u = 1 \quad \text{for } r > R(t)$$

$$r = 0: \quad r^2(\partial T_b/\partial r) = -Q(t) \ \& \ Y_b = 0, \qquad \text{NA}$$

$$r = R(t): \quad T_b = T_f(t) \ \& \ Y_b = 0, \qquad T_u = T_f(t) \ \& \ Y_u = 0$$

$$r \to \infty: \quad \text{NA}, \qquad T_u = 0 \ \& \ Y_u = 1$$

where $Q$ is the heating power of the external source at the center. The flame temperature can be equivalently regarded as a function of flame location. Accordingly, the time derivative of $T_f$ can be determined via chain's rule, $dT_f/dt = U(dT_f/dR)$, where $U = dR/dt$ is the propagation speed of the flame front.

Nevertheless, the preceding formulation is not in closed form since the flame temperature $T_f$ and flame location $R$ remain to be determined. The contribution of chemical reaction to the change of $Y$ and $T$ is characterized by the jump relations at the flame interface. The jump relations across the flame front are derived as the leading order solution of the large activation energy asymptotic analysis (Chen & Ju 2007, Wu & Chen 2012), i.e.,

$$\left(\frac{\partial T_b}{\partial r}\right)_{R^-} - \left(\frac{\partial T_u}{\partial r}\right)_{R^+} = [\epsilon_T + (1-\epsilon_T)T_f]^2 \exp\left[\frac{Z}{2}\frac{T_f - 1}{\epsilon_T + (1-\epsilon_T)T_f}\right] \tag{13}$$

$$\frac{1}{Le_u}\left(\frac{\partial Y_u}{\partial r}\right)_{R^+} - \frac{1}{Le_b}\left(\frac{\partial Y_b}{\partial r}\right)_{R^-} = \left(\frac{\partial T_b}{\partial r}\right)_{R^-} - \left(\frac{\partial T_u}{\partial r}\right)_{R^+} \tag{14}$$

Where the subscripts $R^+$ and $R^-$ denote the corresponding derivatives evaluated at respectively the unburnt and burnt side of the flame front. Substituting the solutions for $T$ and $Y$ into the jump conditions, the desired flame temperature $T_f$ and flame location $R$ could be determined and hence the formulation is in closed form.

## 2.2 Analytical solutions

The time change of the flame front, $R = R(t)$, causes considerable difficulty in solving the governing equations analytically. Mathematically, the flame front can be considered as a moving boundary, which can be removed by introducing a scaled coordinate (Law & Sirignano 1977, Yu & Chen 2020),

$$\sigma_s = \frac{r}{R(t)}, \qquad t_s = \frac{t}{R^2(t)} \tag{15}$$

In terms of $\sigma_s$ and $t_s$, the governing equations become

*Burnt region*

$$\frac{\partial T_b}{\partial t_s} = \frac{\partial^2 T_b}{\partial \sigma_s^2} + \left(\sigma_s RU + \frac{2}{\sigma_s}\right)\frac{\partial T_b}{\partial \sigma_s} \tag{16}$$

$$\frac{\partial Y_b}{\partial t_s} = \frac{1}{Le}\frac{\partial^2 Y_b}{\partial \sigma_s^2} + \left(\sigma_s RU + \frac{1}{Le'}\frac{2}{\sigma_s}\right)\frac{\partial Y_b}{\partial \sigma_s} \tag{17}$$

*Unburnt region*

$$\frac{\partial T_u}{\partial t_s} = \frac{\partial^2 T_u}{\partial \sigma_s^2} + \left(\sigma_s RU + \frac{2}{\sigma_s}\right)\frac{\partial T_u}{\partial \sigma_s} \tag{18}$$

$$\frac{\partial Y_u}{\partial t_s} = \frac{1}{Le}\frac{\partial^2 Y_u}{\partial \sigma_s^2} + \left(\sigma_s RU + \frac{1}{Le}\frac{2}{\sigma_s}\right)\frac{\partial Y_u}{\partial \sigma_s} \tag{19}$$

Because of the differences in boundary conditions, the temperature and mass fraction distributions in burnt and unburnt regions are solved in different ways.

First, we consider the unburnt region. To further simplify the governing equations, we introduce the following pair of F-functions for temperature and mass fraction respectively,

$$F_{uT}(\sigma_s) = \frac{1}{\sigma_s^2}\exp\left[-\frac{1}{2}RU(\sigma_s^2 - 1)\right] \tag{20}$$

$$F_{uY}(\sigma_s) = \frac{1}{\sigma_s^2}\exp\left[-\frac{1}{2}LeRU(\sigma_s^2 - 1)\right] \tag{21}$$

With the help of F-functions, we can define a pair of new coordinates, i.e.,

$$\xi_{uT} = \frac{\int_1^{\sigma_s} F_{uT}(\sigma_s')d\sigma_s'}{\int_1^{\infty} F_{uT}(\sigma_s)d\sigma_s} \tag{22}$$

$$\xi_{uY} = \frac{\int_1^{\sigma_s} F_{uY}(\sigma_s')d\sigma_s'}{\int_1^{\infty} F_{uY}(\sigma_s)d\sigma_s} \tag{23}$$

In terms of $\xi_{uT}$ and $\xi_{uY}$, the governing equations for temperature and mass fraction are simplified to

$$\frac{\partial T_u}{\partial t_s} = \mathcal{F}_{uT}^2 \frac{d^2 T_u}{d\xi_{u,T}^2} \tag{24}$$

$$\frac{\partial Y_u}{\partial t_s} = \frac{\mathcal{F}_{uY}^2}{Le}\frac{d^2 Y_u}{d\xi_{u,Y}^2} \tag{25}$$

where the factors $\mathcal{F}_{uY}$ and $\mathcal{F}_{uT}$ are functions of $\sigma_s$.

$$\mathcal{F}_{uT} = \frac{d\xi_{uT}}{d\sigma_s} = \frac{F_{uT}(\sigma_s)}{\int_1^{\infty} F_{uT}(\sigma_s)d\sigma_s} \tag{26}$$

$$\mathcal{F}_{uY} = \frac{d\xi_{uY}}{d\sigma_s} = \frac{F_{uY}(\sigma_s)}{\int_1^{\infty} F_{uY}(\sigma_s)d\sigma_s} \tag{27}$$

In the $t_s - \xi_{uT}$ and $t_s - \xi_{uY}$ coordinate systems, the initial and boundary conditions become:

$$t_s = 0: \quad T_u = 0; \quad t_s = 0: \quad Y_u = 1$$
$$\xi_{uT} = 0: \quad T_u = T_f; \quad \xi_{uY} = 0: \quad Y_u = 0$$
$$\xi_{uT} = 1: \quad T_u = 0; \quad \xi_{uY} = 1: \quad Y_u = 1$$

The analytical solutions can be obtained as:

$$T_u(\xi_{uT}, t_s) = T_f(1 - \xi_{uT})$$
$$- 2T_f \sum_{n=1}^{\infty} \frac{\sin(n\pi\xi_{uT})}{n\pi} \left( \frac{T_f^0}{T_f} + R^2 U \frac{d \ln T_f}{dR} \frac{e^{\mathcal{F}_{uT}^2 n^2 \pi^2 t_s} - 1}{\mathcal{F}_{uT}^2 n^2 \pi^2} \right) e^{-\mathcal{F}_{uT}^2 n^2 \pi^2 t_s}$$
(28)

$$\approx T_f(1 - \xi_{uT}) - 2T_f^0 \sum_{n=1}^{\infty} \frac{\sin(n\pi\xi_{uT})}{n\pi} e^{-\mathcal{F}_{uT}^2 n^2 \pi^2 t_s}$$

$$Y_u(\xi_{uY}, t_s) = \xi_{uY} + 2 \sum_{n=1}^{\infty} \frac{\sin(n\pi\xi_{uY}) e^{-\mathcal{F}_{uY}^2 n^2 \pi^2 t_s / Le_u}}{n\pi}$$
(29)

where $T_f^0$ refers to the onset flame temperature and will be specified in the subsequent section. During flame propagation, the heat release from reaction and the heat conduction towards the preheat zone tends to balance dynamically. We postulate that the flame temperature $T_f$ does not change rapidly as the flame moving outwardly, i.e.,

$$\frac{d \ln T_f}{dR} \ll 1 \tag{30}$$

Consequently, the approximation in Eq. (28) can be made.

Transforming Eqs. (28) and (29) back to the $r - t$ coordinate system, we obtain the temporal evolution of temperature and mass fraction profiles in the unburnt region. The unsteady solutions for $T_u$ and $Y_u$, given by Eqs. (28) and (29), consist of two components: one is time-independent and characterizes the asymptotic distributions of temperature and mass fraction at the final stage; and the other is time-dependent and represents the change of $T_u$ and $Y_u$ due to heat conduction and mass diffusion. It can be verified that for low to moderate time lapse, the time-dependent component, i.e., the summation of exponential terms, would have comparable magnitude, indicating that the unsteady effect is pronounced during flame kernel development. Therefore, the quasi-steady solution cannot accurately describe the initial development of the flame kernel.

Subsequently, we deal with the burnt region. In the burnt region of $r < R(t)$, the mass fraction of the deficient reactant is always zero, $Y_b = 0$, and so is its gradient at the flame front, i.e. $(\partial Y_b / \partial r)_{R^-} = 0$. Therefore, we only need obtain the analytical solution for temperature in the burnt region.

Without external heating or radiative loss, the temperature in the burnt regime is uniform and equal to the flame temperature $T_f$. The heat addition at the center leads to an increment of temperature from $T_f$, which we denote as $T_b' = T_b - T_f^0$. It satisfies the same governing equation as $T_b$ while the initial condition is replaced by $T_b' = 0$ at $t_s = 0$. To simplify the governing equation, we introduce the radial coordinate weighted temperature discrepancy, $\overline{T}_b = rT_b'$.

$$\frac{\partial \overline{T}_b}{\partial t} = \frac{\partial^2 \overline{T}_b}{\partial r^2} \tag{31}$$

Accordingly, the initial and boundary conditions become

$t = 0$: $\quad \overline{T}_b = 0$ for $r \leq R(t)$

$r = 0$: $\quad \overline{T}_b = Q(t)$

$r = R(t)$: $\quad \overline{T}_b = R(T_f - T_f^0)$

To remove the moving boundary effect due to the flame front propagation, the governing equations for $\overline{T}_b$ can be written in the scaled coordinate $\sigma_s$ and $t_s$ as

$$\frac{\partial \overline{T}_b}{\partial t_s} = \frac{\partial^2 \overline{T}_b}{\partial \sigma_s^2} + \sigma_s RU \frac{\partial \overline{T}_b}{\partial \sigma_s} \tag{32}$$

To simplify the governing equations, we introduce the following pair of F-functions

$$F_{bT}(\sigma_s) = \exp\left(-\frac{1}{2}RU\sigma_s^2\right) \tag{33}$$

With the help of $F_{bT}$, we can define a pair of new coordinates

$$\xi_{bT} = \frac{\int_0^{\sigma_s} F_{b,T}(\sigma_s')d\sigma_s'}{\int_0^1 F_{bT}(\sigma_s)d\sigma_s} = \frac{\text{erf}\left(\sigma_s\sqrt{RU/2}\right)}{\text{erf}\left(\sqrt{RU/2}\right)} \tag{34}$$

In the transformed coordinate $\xi_{bT}$, the governing equation for $\overline{T}_b$ can be written as:

$$\frac{\partial \overline{T}_b}{\partial t_s} = \mathcal{F}_{bT}^2 \frac{d^2 \overline{T}_b}{d\xi_{bT}^2} \tag{35}$$

where

$$\mathcal{F}_{bT} = \frac{d\xi_{bT}}{d\sigma_s} = \frac{2\sqrt{RU/2}\, e^{-\sigma_s^2 RU/2}}{\sqrt{\pi}\,\text{erf}\left(\sqrt{RU/2}\right)} \tag{36}$$

subject to the following initial and boundary conditions

$$t_s = 0: \quad \overline{T}_b = 0$$

$$\xi_{bT} = 0: \quad \overline{T}_b = Q(t)$$

$$\xi_{bT} = 1: \quad \overline{T}_b = R(T_f - T_f^0)$$

The analytical solution for $\overline{T}_b$ can be obtained as

$$\overline{T}_b(\xi_{b,T}, t_s) = Q(t_s) + \xi_{bT}[R(T_f - T_f^0) - Q(t_s)] + 2\sum_{n=1}^{\infty} \sin(n\pi\xi_{bT}) e^{-\mathcal{F}_{bT}^2 n^2 \pi^2 t_s} R_n(t_s) \quad (37)$$

where

$$R_n(t) = -\frac{1}{n\pi}\left[Q(0) + \int_0^t (dQ/d\tau) e^{\mathcal{F}_{bT}^2 n^2 \pi^2 \tau} d\tau\right] \quad (38)$$

The flame temperature $T_f$ and flame radius $R$ can be solved via the matching conditions in Eqs. (13) and (14), which requires the gradients of temperature and mass fraction at the flame front in the physical coordinate. From the chain's rule, the gradients in the unburnt region can be evaluated as

$$\left(\frac{\partial T_u}{\partial r}\right)_{R^+} = -\frac{\hat{\mathcal{F}}_{uT}}{R}\left\{T_f + T_f^0\left[\vartheta_3\left(e^{-\hat{\mathcal{F}}_{uT}^2 \pi^2 t/R^2}\right) - 1\right]\right\} \quad (39)$$

$$\left(\frac{\partial Y_u}{\partial r}\right)_{R^+} = \frac{\hat{\mathcal{F}}_{uY}}{R}\vartheta_3\left(e^{-\pi^2 \hat{\mathcal{F}}_{uY}^2 t/R^2 Le_u}\right) \quad (40)$$

where $\hat{\mathcal{F}}_{uT} = \mathcal{F}_{uT}(\sigma_s = 1)$ and $\hat{\mathcal{F}}_{uY} = \mathcal{F}_{uY}(\sigma_s = 1)$. The Jacobi theta function $\vartheta_3$ denotes the subsequent sum

$$\vartheta_3(x) = 1 + 2\sum_{n=1}^{\infty} x^{n^2} \quad (41)$$

Similarly, the gradients in the burnt region can be written in the following form

$$\left(\frac{\partial T_b}{\partial r}\right)_{R^-} = -\frac{\hat{\mathcal{F}}_{bT}}{R^2}\left\{Q(t_s) + Q(0)\left[\vartheta_4\left(e^{-\hat{\mathcal{F}}_{bT}^2 \pi^2 t/R^2}\right) - 1\right]\right.$$
$$\left. + 2\sum_{n=1}^{\infty}(-1)^n e^{-\hat{\mathcal{F}}_{bT}^2 n^2 \pi^2 t/R^2} \int_0^t (dQ/d\tau) e^{\hat{\mathcal{F}}_{bT}^2 n^2 \pi^2 \tau/R^2} d\tau\right\} \quad (42)$$
$$+ \frac{1}{R}(\hat{\mathcal{F}}_{bT} - 1)(T_f - T_f^0)$$

where $\hat{\mathcal{F}}_{bT} = \mathcal{F}_{bT}(\sigma_s = 1)$, and $\vartheta_4$ is another Jacobi theta function that represents

$$\vartheta_4(x) = 1 + 2\sum_{n=1}^{\infty}(-1)^n x^{n^2} \quad (43)$$

To model the external heating source with finite duration time of $t_h$, we use the Heaviside function $H(t)$ so that external heating is turned on at $t = 0$ and switched off at $t = t_h$, i.e.,

$$Q(t) = Q_m[H(t) - H(t - t_h)] \tag{44}$$

where $Q_m$ represents the magnitude of the heating power. The derivative of $Q(t)$ is given in terms of delta function

$$\frac{dQ}{dt} = Q_m[\delta(t) - \delta(t - t_h)] \tag{45}$$

Therefore, the integral involving $(dQ/d\tau)$ shall be evaluated separately for $t < t_h$ and $t > t_h$

$$\int_0^t (dQ/d\tau) e^{\hat{\mathcal{F}}_{bT}^2 n^2 \pi^2 \tau/R^2} d\tau = \begin{cases} Q_m, & t < t_h \\ Q_m \left(1 - e^{\hat{\mathcal{F}}_{bT}^2 n^2 \pi^2 t_h/R^2}\right), & t > t_h \end{cases} \tag{46}$$

Substituting Eq. (46) into (42) yields

$$\left(\frac{\partial T_b}{\partial r}\right)_{R^-} = \frac{\hat{\mathcal{F}}_{bT}}{R}(T_f - T_f^0) - \frac{Q_m}{R^2}\hat{\mathcal{F}}_{bT} S(t, U, R) \tag{47}$$

where the function $S$ is defined as

$$S(t, U, R) = \begin{cases} \vartheta_4\left(e^{-\hat{\mathcal{F}}_{bT}^2 \pi^2 t/R^2}\right), & t < t_h \\ \vartheta_4\left(e^{-\hat{\mathcal{F}}_{bT}^2 \pi^2 t/R^2}\right) - \vartheta_4\left(e^{-\hat{\mathcal{F}}_{bT}^2 \pi^2 (t-t_h)/R^2}\right), & t > t_h \end{cases} \tag{48}$$

Substituting Eqs. (39), (40), (42) and $(\partial Y_b/\partial r)_{R^-} = 0$ into Eqs. (13) and (14), one obtains the following expression for flame temperature

$$T_f = T_f^0 + \frac{\hat{\mathcal{F}}_{uY}\vartheta_3\left(e^{-\pi^2 \hat{\mathcal{F}}_{uY}^2 t/R^2 Le_u}\right)/Le_u + Q_m \hat{\mathcal{F}}_{bT} S(t, U, R)/R - \hat{\mathcal{F}}_{uT} T_f^0 \vartheta_3\left(e^{-\hat{\mathcal{F}}_{uT}^2 \pi^2 t/R^2}\right)}{\hat{\mathcal{F}}_{bT} + \hat{\mathcal{F}}_{uT} - 1} \tag{49}$$

and the following condition characterizing the consumption of reactant by chemical reaction,

$$\frac{\hat{\mathcal{F}}_{uY}}{Le_u R}\vartheta_3\left(e^{-\pi^2 \hat{\mathcal{F}}_{uY}^2 t/R^2 Le_u}\right) = [\epsilon_T + (1-\epsilon_T)T_f]^2 \exp\left[\frac{Z}{2}\frac{T_f - 1}{\epsilon_T + (1-\epsilon_T)T_f}\right] \tag{50}$$

At the initial instant, $t = 0$, the Jacobi theta functions in Eq. (50) is equal to $\vartheta_3(1)$, which is infinitely large, while the chemical reaction rate has a finite value. Such inconsistency in Eq. (50) implies that the flame kernel cannot be established at $t = 0$. As time $t$ increases, the Jacobi theta function decreases rapidly, and at some instant, denoted by $t = t_{ig}$, the left- and right hands of Eq. (50) becomes identical, indicating the appearance of a flame kernel, which progressively accelerates from $U = 0$. According to the definition of $T_f^0$, i.e., $T_f^0 = T_f(t = t_{ig})$, the onset flame temperature can be determined through Eq. (49), giving

$$T_f^0 = \frac{1}{Le} \frac{\vartheta_3\left(e^{-\pi^2 t_{ig}/R_0^2 Leu}\right)}{\vartheta_3\left(e^{-\pi^2 t_{ig}/R_0^2}\right)} + \frac{Q_m}{R_0} \frac{S(t_{ig}, 0, R_0)}{\vartheta_3\left(e^{-\pi^2 t_{ig}/R_0^2}\right)} \tag{51}$$

The factors $\hat{\mathcal{F}}_{uT}$, $\hat{\mathcal{F}}_{uY}$, and $\hat{\mathcal{F}}_{bT}$ are functions of flame location $R$ and propagating speed $U = dR/dt$. For $U = 0$, we have $\hat{\mathcal{F}}_{uT} = \hat{\mathcal{F}}_{uY} = \hat{\mathcal{F}}_{bT} = 1$.

Successful ignition refers to the generation of a self-sustained expanding flame. In the absence of external heating, there is a critical radius, below which the heat loss by conduction dominates over the heat release from chemical reaction, and thereby successful ignition cannot occur (Joulin 1985, Chen & Ju 2007). When the mixture's Lewis number is not considerably greater than unity, the critical radius is identical to the flame ball radius (He 2000, Chen et al. 2011). Setting $U = 0$ in the matching conditions (49) and (50), the flame ball radius can be obtained as

$$R_Z = \frac{Le}{[1 + (Le-1)\epsilon_T]^2} \exp\left[\frac{Z}{2} \frac{Le - 1}{1 + (Le - 1)\epsilon_T}\right] \tag{52}$$

which is a function of mixture's thermophysical properties. The above expression for flame ball radius agrees with Zel'dovich theory (Ya et al. 1980). The flame ball radius becomes larger at higher Lewis number.

Substituting Eq. (49) into Eq. (50), one obtains an implicit ordinary differential equation for flame raidus $R$ subject to the initial conditions $R = R_0$ at $t = t_{ig}$. When $R = R(t)$ is obtained, the flame propagation speed is obtained via $U(t) = dR/dt$. Substituting $R(t)$ and $U(t)$ into Eq. (49), the flame temperature is obtained and thereby the flame kernel development is completely solved. Then, the unsteady evolution of temperature and mass fraction distributions during flame ignition process can be obtained from Eqs. (28), (29) and (43). Note that in the burnt region we always have $Y_b = 0$.

In previous studies considering flame kernel evolution, the quasi-steady approximation has been widely adopted. In quasi-steady theory, the time derivatives in the conservation equations are neglected and the following expressions can be obtained from the matching conditions (He 2000, Chen & Ju 2007):

$$T_f = \left[\frac{1}{Le} \frac{e^{-UR(Le-1)}}{\int_R^\infty \tau^{-2} e^{-U\tau Le} d\tau} + Q_m\right] \int_R^\infty \tau^{-2} e^{-U\tau} d\tau \tag{53}$$

$$\frac{e^{-UR}}{R^2} \left(\frac{T_f}{\int_R^\infty \tau^{-2} e^{-U\tau} d\tau} - Q_m\right) = [\epsilon_T + (1 - \epsilon_T)T_f]^2 \exp\left[\frac{Z}{2} \frac{T_f - 1}{\epsilon_T + (1 - \epsilon_T)T_f}\right] \tag{54}$$

Substituting Eq. (53) into (54) yields a nonlinear equation which describes the change of flame propagation speed with flame radius during the flame kernel development, i.e., $U = U(R)$. In particular, the flame ball radius $R'_Z$ can be determined in quasi-steady theory by setting $U = 0$ in Eqs. (53) and (54), and it is identical with that derived based on transient formulation given by Eq. (52). When the flame radius becomes infinitely large, the planar flame solution is reached, and both the non-dimensional flame propagation speed and flame temperature are unity. At such conditions, the matching conditions for both quasi-steady theory and transient formulation become identical again. Therefore, in the limits of both stationary flame ball ($U = 0$) and planar flame ($R \to \infty$), the present transient formulation degenerates to the quasi-steady theory.

The transient formulation in this study rigorously takes into account the unsteady evolution of temperature and mass fraction distributions during flame ignition process. The analytical solutions explicitly indicate that the temperature and mass fraction profiles on each side of the flame front change with time. However, in quasi-steady theory, the temperature and mass fraction are regarded as functions of spatial coordinate, characterizing their distributions in the final state, i.e., subsequent to sufficient long-term evolution. According to the matching conditions at the flame interface, Eqs. (13) and (14), transience in temperature and mass fraction gradients implies time-dependence of flame temperature, which directly affects the dynamical behavior of the expanding flame. Moreover, finite energy deposition can be appropriately considered in the present transient formulation. The heating duration introduces a definite characteristic time, which is not considered in the quasi-steady theory.

## 3. Results and Discussion

The transient propagation of the ignition kernel and the critical heating power and MIE for successful ignition can be described by Eqs. (49) and (50). Using these equations, we shall demonstrate how the unsteady effect influences the flame initiation process and critical ignition conditions. For typical premixed flames, we choose $Z = 10$ and $\epsilon_T = 0.15$ according to previous studies (Chen et al. 2011, Wu & Chen 2012).

### 3.1 Flame initiation without central heating

We first consider the case without ignition energy deposition at the center, i.e., $Q_m = 0$. The flame kernel development can be described in the U-R diagram as shown in Fig. 1. The dynamic behavior of flame front propagation based on the quasi-steady theory agrees qualitatively with that predicted by the transient formulation. Specifically, Fig. 1 shows that the transient formulation and quasi-steady theory yield identical flame ball radii at $U = 0$ and consistently interpret the flame kernel propagation toward quasi-planar flame at $R \to \infty$. Moreover, the transient formulation reconciles with the quasi-steady theory in terms of the Lewis number effect. Specifically, for mixture with Lewis number close to or smaller than unity, the flame ball radius is the critical radius beyond which the flame can propagate outwardly in self-sustained manner. When the Lewis number is sufficiently small, e.g., $Le = 0.5$, the curvature effect creates a super-adiabatic condition, driven by which the flame kernel accelerates rapidly with propagation speed considerably higher than that of planar flame. This implies that a flame can be ignited beyond flammability limit and undergoes self-extinguishing at certain conditions (Ronney 1989, Ronney & Sivashinsky 1989). The curvature effect tends to be alleviated when the flame propagates outwardly.

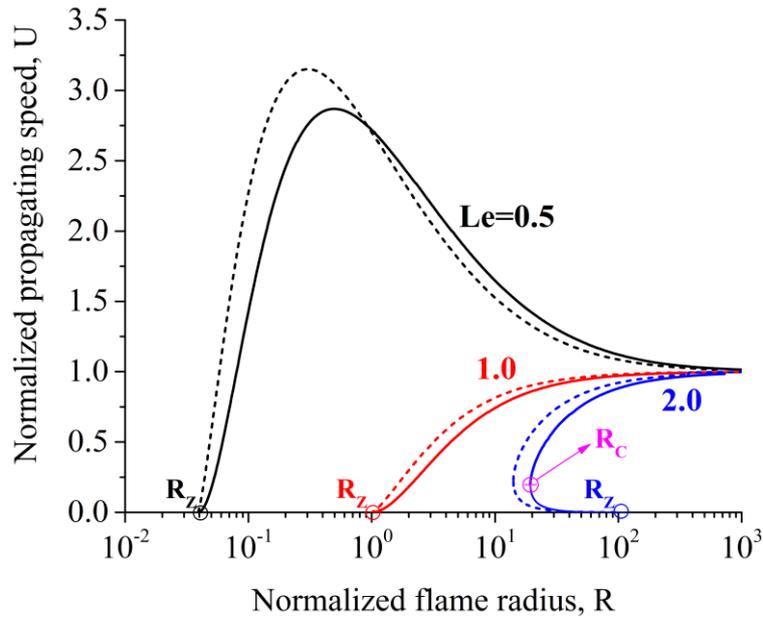

Figure 1. Change of flame propagation speed with flame radius for different Lewis numbers. The solid and dashed lines represent results from transition formulation and quasi-steady theory, respectively. The mixture's Lewis number ( $Le = 0.5, 1.0, 2.0$ ) is indicated by the color (black/red/blue) of lines. $R_Z$ and $R_C$ respectively denote the flame ball radius and critical radius for flame initiation.

When the Lewis number is higher than some critical value slightly above unity, e.g., $Le = 2.0$, the U-R diagram exhibits a C-shaped curve. The turning point of the C-shaped U-R curve corresponds to the critical radius of $R_c$. The flame kernel structure cannot be established for $R < R_c$ due to severe conductive heat loss in large curvature condition. Figure 1 shows that for $Le = 2$, the flame ball radius is larger than the critical radius, i.e., $R_Z > R_c$. Therefore, the stationary flame ball radius is no longer the minimum radius that controls flame initiation in mixtures with large Lewis number (Chen et al. 2011).

From quantitative aspect, results in Fig. 1 indicate that the U-R relation predicted by the quasi-steady theory deviates from that from solutions of transient formulation for intermediate values of $U$ and $R$. Such discrepancy could be elucidated by examining the time scales characterizing the flame propagation and change of temperature gradient. The reference time scale for flame propagation can be defined as $t_{\text{fp}} = 1/U$. According to the unsteady solution of $\partial T_u/\partial t$ given by Eq. (39), the characteristic time for the change of temperature gradient is $t_{\text{un}} = R^2 \mathcal{F}_{uT}^2/\pi^2$.

For low to moderate values of $RU$, the factors $\mathcal{F}_{uT}$ and $\mathcal{F}_{bT}$ can be expanded in series of $RU$, where the first-order correction must be retained, yielding

$$\mathcal{F}_{uT} \approx 1 - \sqrt{\frac{\pi R U}{2}}, \qquad \mathcal{F}_{uY} \approx 1 - \sqrt{\frac{Le \pi R U}{2}} \tag{55}$$

Substituting the simplified $\mathcal{F}_{uT}$ and $\mathcal{F}_{uY}$ into the matching conditions, the flame propagation speed can be estimated by

$$U \approx \frac{2}{\pi R Le} \frac{(R_Z/R - 1)^2}{(R_Z/R + F_R)^2} \tag{56}$$

where the factor $F_R$ is

$$F_R = \frac{(1 - 1/\sqrt{Le})\{4(\epsilon_T - 1)[1 + \epsilon_T(Le - 1)] - LeZ\}}{2[1 + (Le - 1)\epsilon_T]^2} \tag{57}$$

According to the definition of reference time for flame propagation and Eq. (56), $t_{\text{fp}}$ is given by

$$t_{\text{fp}} \sim \frac{\pi R Le}{2} \frac{(R_Z/R + F_R)^2}{(R_Z/R - 1)^2} \tag{58}$$

At the onset of flame kernel, the radius is close to that of flame ball, i.e., $R_Z/R - 1 \ll 1$, and thereby $t_{\text{fp}} \gg 1$. However, the characteristic time for temperature gradient evolvement, according to its definition, appears at most of order unity, i.e., $t_{\text{un}} \sim O(1)$. The exceedingly slow propagating speed provides sufficient time for the local temperature gradient to develop into the steady-state distribution. Therefore, for flame radius close to flame ball size, the unsteady effect is negligible, resulting in the consistency between the quasi-steady theory and transient formulation in the limit of $U \to 0$.

At moderate values of $RU$, one has

$$t_{\text{fp}}/t_{\text{un}} \sim \pi R Le (R_Z/R + F_R)^2 / 2(R_Z/R - 1)^2 \sim O(1) \tag{59}$$

It indicates that the flame propagation speed and the time change rate of local temperature gradient would be of the same order of magnitude, implying that the unsteady effects may have consequential impacts on flame propagation.

As flame continues to propagate outwardly, we have $R \gg 1$ and $U \approx 1$. The factor $\mathcal{F}_{uT}$ can be expanded by treating $1/(RU)$ as a small parameter, yielding $\mathcal{F}_{uT} \approx 1/R$. The characteristic time for temperature gradient evolution can be estimated by $t_{\text{un}} = 1/\pi^2$. Meanwhile, the flame propagation time is given by $t_{\text{fp}} = 1/U \approx 1$. Therefore, we have $t_{\text{fp}} \approx 10 t_{\text{un}}$. This indicates that when the expanding flame is approaching the quasi-planar flame, the local temperature gradient has sufficient time to develop into the steady state distribution. This again leads to the consistency between quasi-steady theory and transient formulation in the limit of $R \to \infty$.

The above time scale analysis indicates that for low to moderate Lewis numbers, the unsteady effects becomes important at some intermediate flame radius, where the propagating speed is comparable with the time change rate of local temperature gradient. Figure 2 plots the profiles of the temperature and mass fraction of the deficient reactant for a propagating flame with $Le = 1$ at the instant when the flame radius is $R = 2$. It shows that in the unburnt region ($r > 2$), the gradients of temperature and mass fraction predicted by the quasi-steady theory are slightly smaller than those considering the unsteady effects because of long-time evolution. However, the quasi-steady theory and transient formulation predict almost identical flame temperature. In the quasi-steady theory, the diffusion rate of reactant towards the flame front tends to be lower due to the relatively low mass fraction gradient in the unburnt region. Therefore, an additional amount of reactant must be supplied by enhanced convection, which accounts for the slight elevation of U-R diagram for quasi-steady theory as shown in Fig 1.

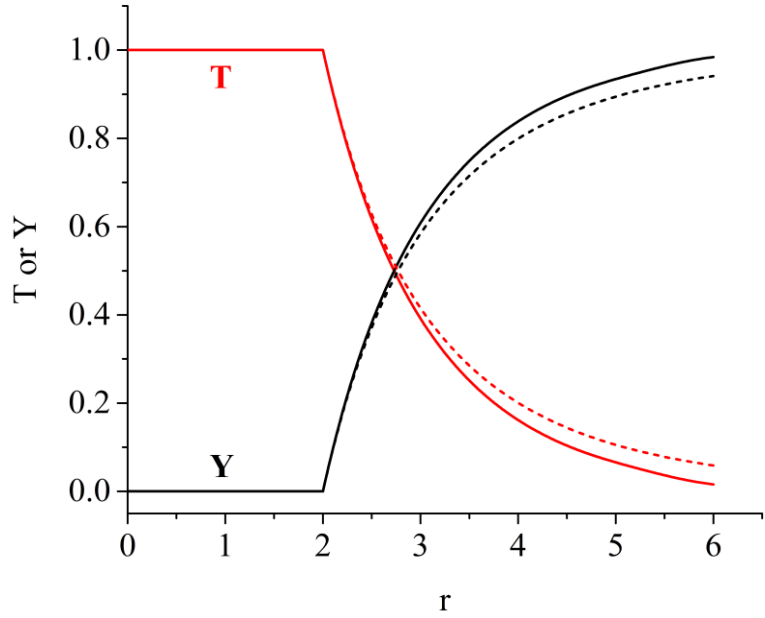

Figure 2. Profiles of temperature and mass fraction of the deficient reactant determined by transient formulation (solid lines) and quasi-steady theory (dashed lines) for $R = 2$, $Le = 1$ and $Q_\mathrm{m} = 0$.

As mentioned before, in the final stage for the expanding flame approaching the planar flame, it has $t_\mathrm{fp}/t_\mathrm{un} \sim O(10)$, and thereby the unsteady effect tends to be negligible. For mixtures with relatively large Lewis number, the stationary flame ball radius differs from the critical radius characterizing flame ignition. Figure 1 shows that the critical radius predicted by quasi-steady theory, $R_C = 14$, is shorter than that based on transient formulation, $R_C = 19$, while the critical speeds at the turning point of the $C$-shaped U-R diagram are almost identical. Specifically, the product $R_c U_c$ for $Le = 2$ has a moderate magnitude, which implies that the unsteady effect would become important according to Eq. (59). According to Eq. (39), the temperature gradient ahead of the spherical flame is proportional to the inverse of the flame radius, i.e., $(\partial T_u/\partial r)_{r=R} \sim 1/R$, and gradually decays as spherical flame expanding. The critical radius defines a particular magnitude of $(\partial T_u/\partial r)_{r=R}$ beyond which the flame structure cannot be established due to excessive heat loss in the preheat zone (Deshaies & Joulin 1984, Chen & Ju 2007). Figure 2 shows that the temperature and mass fraction profiles based on transient formulation are steeper than those given by quasi-steady theory in the unburnt region. Accordingly, the transient formulation yields a larger critical radius to relax the local temperature and mass fraction gradients in the preheat zone to ensure the successful establishment of spherical flame structure. A detailed calculation of critical radius at various Lewis numbers will be presented in the next sub-section.

## 3.2 Flame initiation with constant central heating

In the transient ignition model introduced in Section 2, central heating via the boundary condition at $r = 0$, i.e., $r^2(\partial T_b/\partial r) = -Q(t)$, is used to mimic the ignition energy deposition. We first consider the simplified case of constant central heating, i.e., $t_h \to \infty$, which was considered in previous quasi-steady analysis on ignition (He 2000, Chen et al. 2011). In practice, the duration of the ignition energy deposition is limited. We shall consider the case of finite-duration central heating in the next sub-section.

The central heating results in a high temperature region and generates an ignition kernel with small radius. Figure 3 shows the distributions of the temperature and mass fraction of the deficient reactant for a propagating spherical flame with the radius of $R = 2$, which is induced by the constant central heating of $Q_m = 0.1$. It is seen that central heating leads to a significant increment in temperature around the center and it continuously supplies energy toward the flame front. The temperature profiles in the unburnt regions predicted by quasi-steady theory tends to be more temperate than that based on transient formulation due to the fundamental characteristic of long-term evolution in quasi-steady theory. Correspondingly, the flame propagation speed determined by quasi-steady theory is higher than that based on transient formulation according to discussion in the preceding sub-section without central heating. In quasi-steady theory, the temperature distribution in the burnt region is determined by Chen and Ju (2007)

$$T_b(r) = T_f + Q_m \int_r^R \frac{e^{-U\tau}}{\tau^2} d\tau \tag{60}$$

which tends to be increasingly flat close to the flame front as propagation speed becomes higher. Accordingly, the temperature in the burnt region predicted by the quasi-steady theory appears to be slightly lower than that determined by transient formulation. Nevertheless, the central heating plays a dominant role in affecting the temperature profiles in the burnt region, rendering the unsteady effects to be secondary.

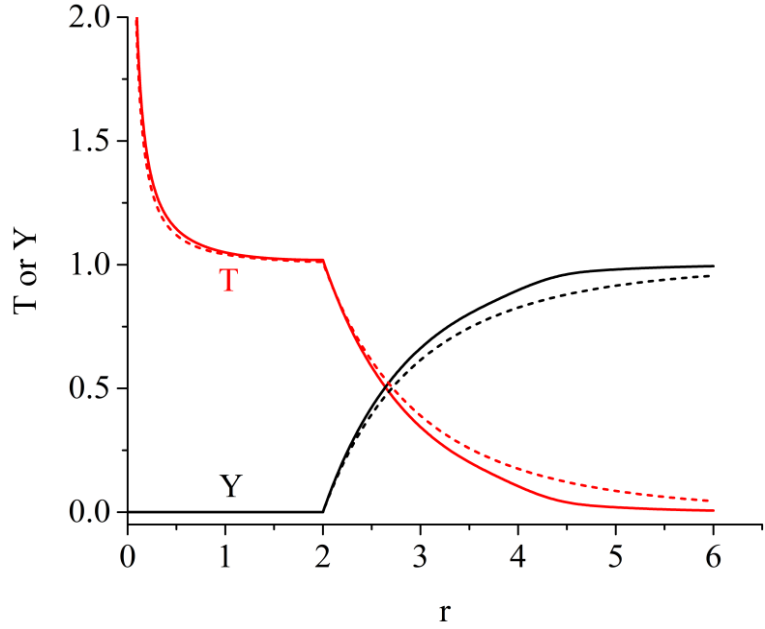

Figure 3. Profiles of temperature and mass fraction of the deficient reactant determined by transient formulation (solid lines) and quasi-steady theory (dashed lines) for $R = 2$, $Le = 1$ and $Q_m = 0.1$.

Figure 4 shows the U-R diagrams for different heating powers at $Le = 1$ and $Le = 2$. The external heating reduces the critical radius, i.e., $R_Z^+$ and $R_c$ for successful flame initiation. Meanwhile, relatively low heating power leads to the emergence of an inner flame ball solution with radius $R_Z^-$, e.g., situations with $Q_m = 0.05$ at $Le = 1$ and $Q_m = 1.0$ at $Le = 2$ as shown in Fig 4. The inner flame ball is stable (Champion et al. 1986, Clavin & Searby 2016). Therefore, for external heating power less than a critical value, denoted by $Q_{cr}$, the flame kernel ignited nearby the heating source is trapped within the inner flame ball instead of continuously propagating outwardly. This means that ignition fails for $Q_m < Q_{cr}$. The minimum distance between U-R diagrams characterizing the development of inner flame and outwardly propagating spherical flames is denoted by $\Lambda$, which decreases with the heating power and identically vanishes at the critical power. For $Q_m = 0.07$ at $Le = 1$ and $Q_m = 2.5$ at $Le = 2$ (i.e., the blue solid lines in Fig. 4), the U-R diagram becomes a continuous curve originating from point O at $R = R_0$ and $U = 0$ (where flame kernel is ignited due to energy deposition) to point D with $R \to \infty$ and $U = 1$ (where planar flame structure is established). The flame kernel can propagate outwardly along this curve, denoted by OABCD, indicating that successful ignition is achieved for $Q_m > Q_{cr}$.

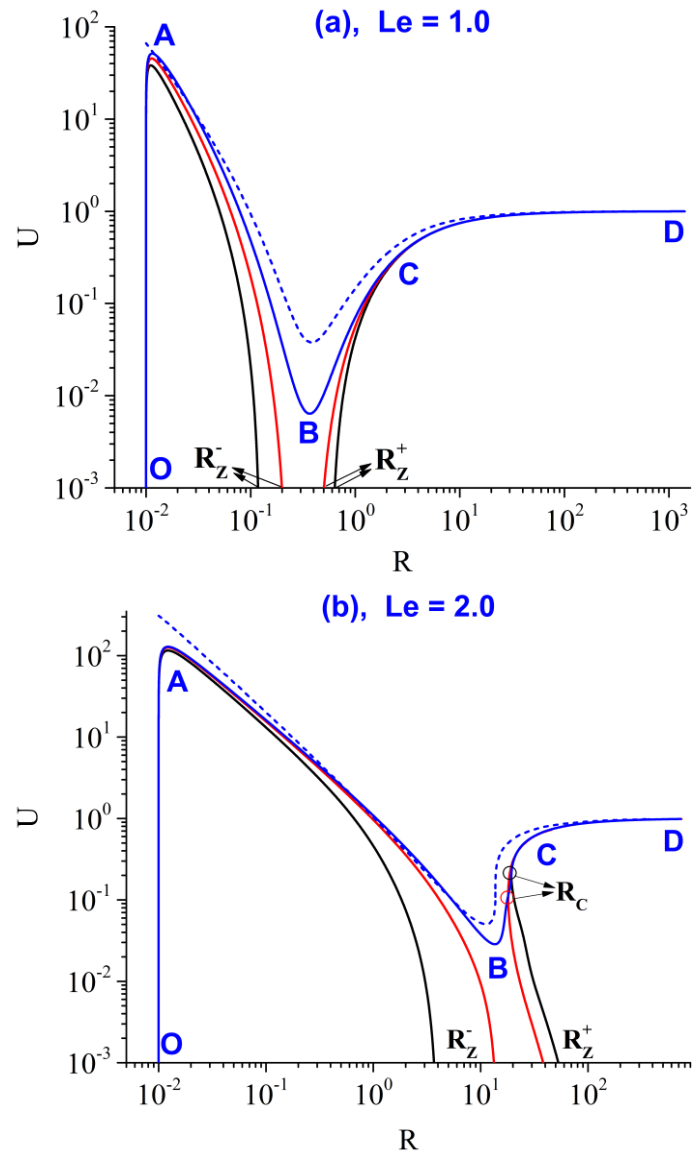

Figure 4. Change of flame propagation speed with flame radius for different central heating powers. The solid lines are solutions from the transient formulation, while the dashed lines are results from quasi-steady theory. The heating powers are indicated by colors of the solid/dashed lines. (a) black for $Q_m = 0.05$, red for $Q_m = 0.06$, and blue for $Q_m = 0.07$; (b) black for $Q_m = 1.0$, red for $Q_m = 2.0$, and blue for $Q_m = 2.5$. $R_Z$ and $R_C$ respectively denote the flame ball radius and critical radius for flame initiation.

Subject to central heating, successful flame initiation comprises four stages: (I), fast establishment of the ignition kernel (curve OA in Fig. 4); (II), ignition-energy-supported flame kernel propagation (curve AB); and (III), unsteady transition of the flame kernel (curve BC); and (IV), quasi-steady spherical flame propagation before its transition to a planar flame (curve CD). In stage I, energy deposition via central heating provides a local high temperature environment, which leads to the ignition of the reactive premixture and the appearance of the ignition kernel. Usually, external heating is highly concentrated, implying that the ignition kernel would be very restricted in spatial dimension. According to our calculation, the qualitative behavior of U-R diagram in the flame-kernel-establishing stage is quite insensitive to the change of onset flame radius $R_0$. Here we set $R_0 = 0.01$ when evaluating the impacts of other affecting parameters.

The temporal evolution of temperature profiles at each stage is shown in Fig 5 for $Le = 2.0$ and $Q_m = 2.5$. Since the onset flame radius $R_0$ is exceedingly small, the temperature profiles nearby, shown in Fig. 5(a), are presented in zoom-in perspective. Large flame curvature characterized by $1/R_0$ results in high temperature gradients on both sides, whose difference is attributed to the heat release from chemical reaction. Both central heating and exothermic chemical reaction give rise to temperature increase inside of the flame kernel, which further facilitates the temperature-sensitive chemical reaction. Such positive feedback leads to ignition kernel acceleration until it achieves the maximum propagation speed (see point A in Fig. 4), at which the heat generation by chemical reaction and heat supply from central heating are balanced with the heat loss by conduction in the preheat zone at the flame front. Consequently, the ignition kernel is fully established.

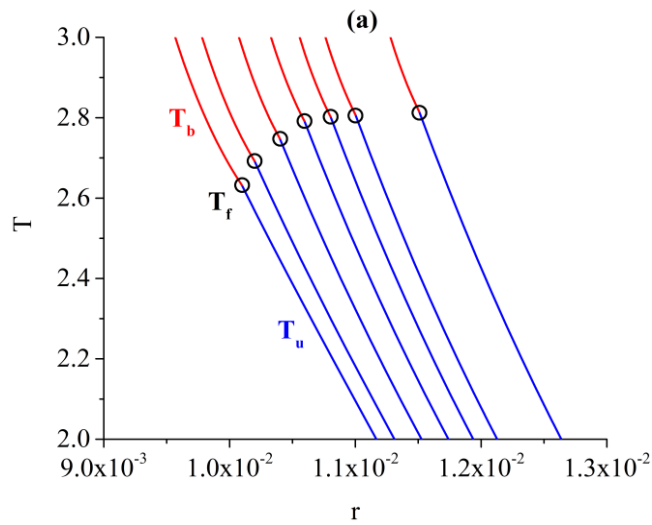

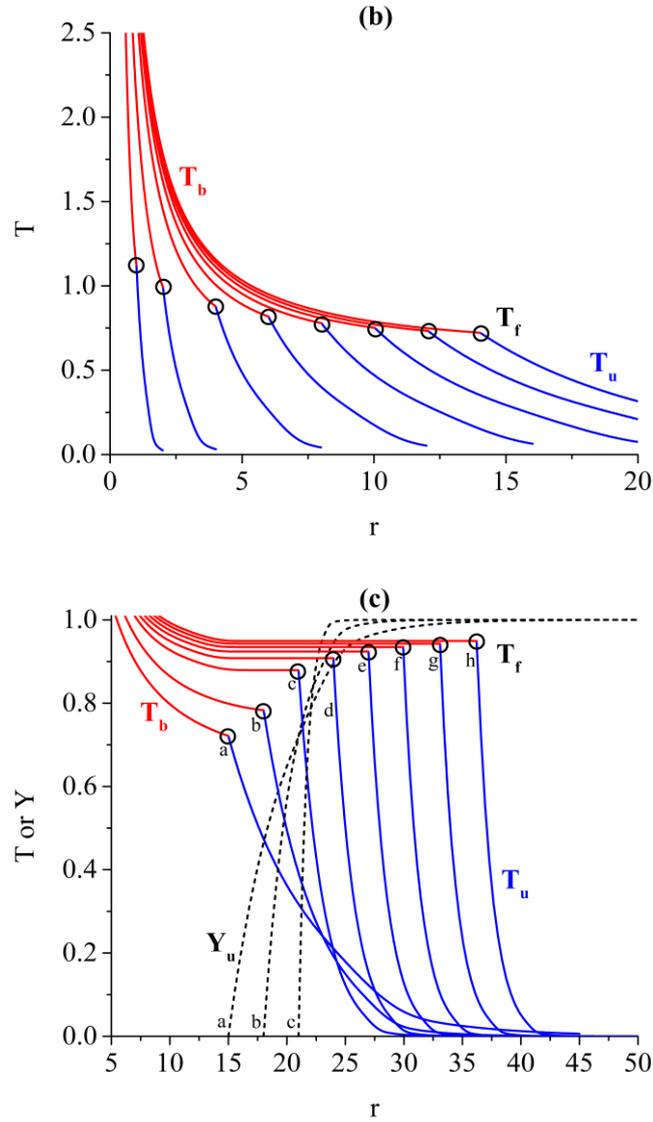

Figure 5. Temporal evolution of the temperature distributions during different ignition stages: (a), stage I for fast establishment of the ignition kernel; (b), stage II for ignition-energy-supported flame kernel propagation; and (c), stage III for unsteady transition of the flame kernel and stage IV for quasi-steady spherical flame propagation. The Lewis number is $Le = 2.0$ and the central heating power is $Q_m = 2.5$. The circles represent flame temperature and flame radius. The distributions for the mass fraction of the deficient reactant are also shown in Fig. (c).

Figure 4 shows that after achieving the maximum propagation speed at point A, the flame kernel continuously decelerates along curve AB. Accordingly, Fig. 5(b) shows that the flame temperature gradually decays in stage II of ignition-energy-supported flame kernel propagation. During this stage, the temperature gradient on the burnt side of the flame kernel is still negative, indicating that heat from the central energy deposition is supplied to the flame front and thereby the flame kernel propagation is still supported by central heating. During the flame kernel propagation in stage II, the flame radius becomes larger and the heat supplied to the flame front by central ignition becomes smaller, and thereby both flame temperature and flame propagation speed become lower. When the heating power is below the critical value ($Q_m < Q_{cr}$), the flame propagation speed eventually reduces to zero in stage II, approaching the inner flame ball solution and resulting ignition failure (see the left branch of U-R curve in Fig. 4 for $Q_m = 0.06$ at $Le = 1$ and $Q_m = 2.0$ at $Le = 2$).

Under supercritical heating (i.e., $Q_m > Q_{cr}$), the flame kernel is capable to pass the critical radius with positive propagation speed at the end of developing stage (around point B in Fig. 4). Then the flame kernel continuously propagates outwardly along curve BC in Fig. 4. The evolution of the temperature profiles during the unsteady transition stage is shown in Fig 5(c). The flame temperature starts to increase again. From energy conservation perspective, it can be inferred that the competition between heat generation via chemical reaction and heat loss via conduction to the preheat zone is responsible for flame temperature increasing. The intensified chemical reaction requires more reactant premixture to be transported towards flame front, which is revealed by the steepening of mass fraction gradients in the preheat zone. The increasing flux of fresh reactant mixture comparatively lowers the temperature ahead of the flame front, as indicated in Fig 5(c). When the spherical flame propagates in a quasi-steady manner, a stable balance between heat release from chemical reaction and heat conduction to warm up the reactant premixture is achieved, and the flame is affected by continuously decaying stretch rate.

Under subcritical heating (i.e., $Q_m < Q_{cr}$) and for Lewis number not considerably larger than unity, the U-R diagram has two branches (see the black and red lines in Fig. 4a): the left branch for the formation of the ignition-energy-supported, stationary flame ball with radius of $R_Z^-$, and the right branching for the continuous expansion of spherical flame originating from the conventional flame ball with radius of $R_Z^+$. With the increase of the heating power, Fig. 6 shows that $R_Z^-$ quickly increases while $R_Z^+$ slightly decrease. At the critical heating condition, $R_Z^-$ and $R_Z^+$ merges. Therefore, the critical or minimum ignition power is determined by the requirement of $R_Z^- = R_Z^+$. For mixtures with relatively large Lewis numbers, Fig. 4(b) shows that the critical radius $R_c$ is shorter than outer flame ball radius $R_Z^+$. Consequently, Fig. 6 shows that for $Le = 1.5$ and 2, successful flame initiation, characterized by the vanishing of the minimum distance between two branches of U-R diagrams respectively describing the inner flame ball and outwardly propagating spherical flame, occurs prior to the merging of $R_Z^-$ and $R_Z^+$.

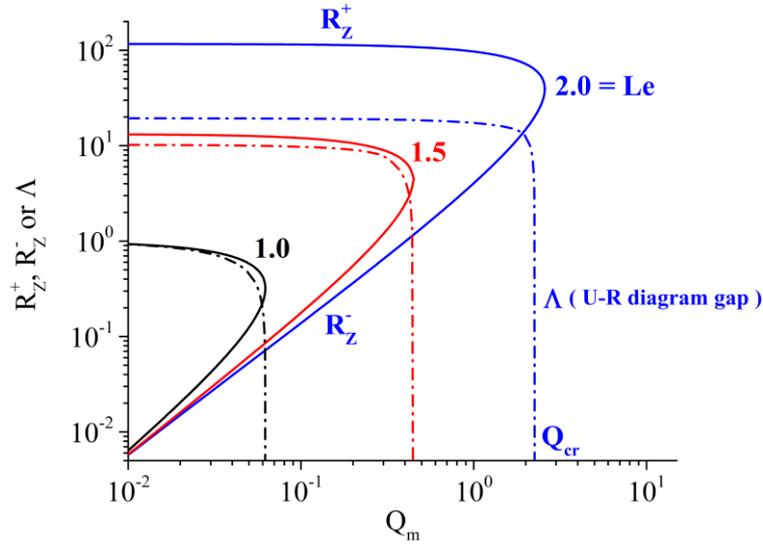

Figure 6. The change of flame ball radii, $R_Z^-$ and $R_Z^+$, and the minimum distance between two branches of U-R diagrams, $\Lambda$, with the central heating power for different Lewis numbers of $Le = 1, 1.5,$ and $2$.

To show the effect of Lewis number on the critical ignition conditions, we calculate the critical heating power and critical ignition radius for different Lewis numbers. The results are depicted in Fig. 7, in which the data from the transient formulation and quasi-steady theory are shown together for comparison. Both the critical heating power and critical ignition radius are shown to increase monotonically with the Lewis number, which is consistent with previous results (Chen et al. 2011).

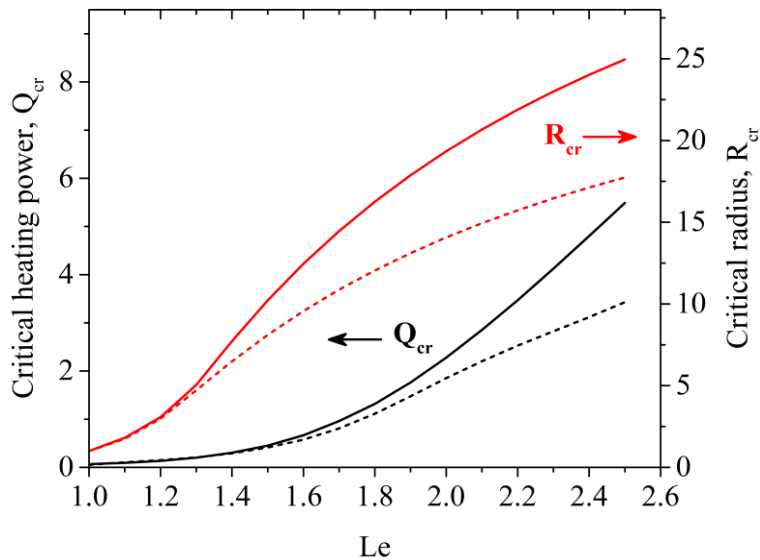

Figure 7. Change of critical heating power and critical ignition radius with the Lewis number. The solid lines are solutions from the transient formulation, while the dashed lines are results from quasi-steady theory.

Figure 7 shows that the critical heating power and critical ignition radius predicted by the transient formulation are higher than those by the quasi-steady theory. The critical radius for ignition is characterized by the maximum conductive heat loss in the preheat zone that can support flame structure. As indicated by results in Figs. 2 and 3, the local temperature and mass fraction distributions ahead of the flame front predicted by the transient formulation are steeper than those predicted by quasi-steady theory, implying a more intensive conductive heat loss and slower flame propagation speed (as shown in Fig. 1) in transient formulation at the same flame radius. Therefore, the critical radius determined by transient formulation is larger than that based on quasi-stead theory to ensure the establishment of flame kernel structure. This indicates that when unsteady effect is taken into account, more intensive energy deposition is required to overcome the flame deceleration during the flame kernel propagation in stage II and thus to ensure successful flame initiation. Consequently, the critical heating power determined by transient formulation tends to be increasingly greater than that based on quasi-steady theory.

Figure 8 shows the scaling relation between the critical heating power and the critical radius. In the quasi-steady theory, the predicted critical heating powers (represented by the red symbols in Fig. 8) appear to change linearly with the cube of the critical radius, which is consistent with previous studies (Chen et al. 2011). However, the critical heating powers determined by transient formulation (see the black symbols in Fig. 8), change more rapidly than the cube of the critical radius and they satisfies a modified scaling relation, $Q_{cr} \sim R_{cr}^{3+\delta}$ with $\delta > 0$.

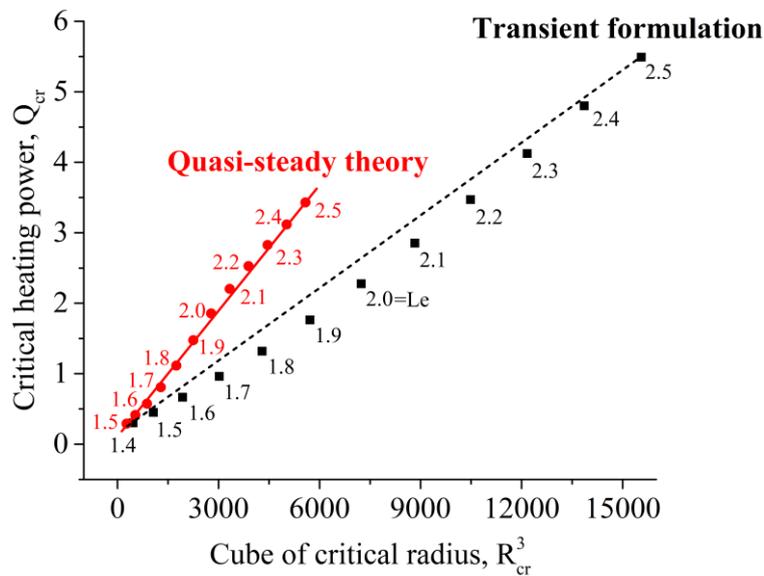

Figure 8 Change of critical heating power with the cube of critical ignition radius. The symbols

represent results from quasi-steady theory or transient formulation, and the lines represent the scaling relationship of $Q_{cr} \sim R_{cr}^3$.

The external heating creates a thermal conduction channel from the center to the flame kernel. In the quasi-steady theory, the net energy transfer rate through each element of unit volume within the flame kernel is inversely proportional to the total volume of the flame kernel, i.e., $Q_m/R^3$. Depending on the differential diffusion characteristics of the reactant mixture, there exists a maximum heat conduction rate in the preheat zone, denoted by $h_{cr}$, beyond which the flame structure cannot be established. The critical radius for flame initiation can be determined with the knowledge of $h_{cr}$ as a function of Lewis number. Moreover, the heat release rate due to chemical reaction at the flame front, $q$, reveals the exothermicity of the reactant mixture and thus is independent of critical radius. Therefore, under critical heating situation the balance of energy flux at the critical radius can be written as follows

$$\frac{Q_{cr}}{R_{cr}^3} + q \sim h_{cr} \tag{61}$$

which qualitatively explains the linearly relation between $Q_{cr}$ and $R_{cr}^3$.

Considering transient evolution of temperature inside the flame kernel, the thermal conduction from heating source to the flame front is associated with more rapidly temperature rising close to the flame kernel center, and accordingly the net energy flux by thermal conduction decays radially. Since the critical radius for $Le > 1$ are larger than unity, phenomenologically, the energy balance at the critical radius tends to be modified as

$$\frac{Q_{cr}}{R_{cr}^{3+\delta}} + q \sim h_{cr} \tag{62}$$

where the modelling factor $\delta$ is greater than zero and underlines the reduction of thermal conduction rate at the flame front in comparison with quasi-steady theory. Arranging Eq. (62) gives that $Q_{cr} \sim R_{cr}^{3+\delta}$, which coincides with the downward-convex distribution of critical heating power (black squares) with the cube of critical radius as shown in Fig. 8. The evaluation of the modelling factor $\delta$ involves rigorously dealing with the transient temperature distribution in the burnt region during flame kernel development, which appears to be an exceedingly complicated task for analytical treatment and is beyond the scope of this study.

## 3.3 Flame initiation with finite duration heating

In this sub-section, we consider the ignition induced by finite-duration central heating, which is closer to practical ignition than the constant central heating. For duration time of $t_h$ and heating power of $Q_m$, the ignition energy is $E_{ig} = Qt_h$. Figure 9 shows the U-R diagram for different heating power and duration time.

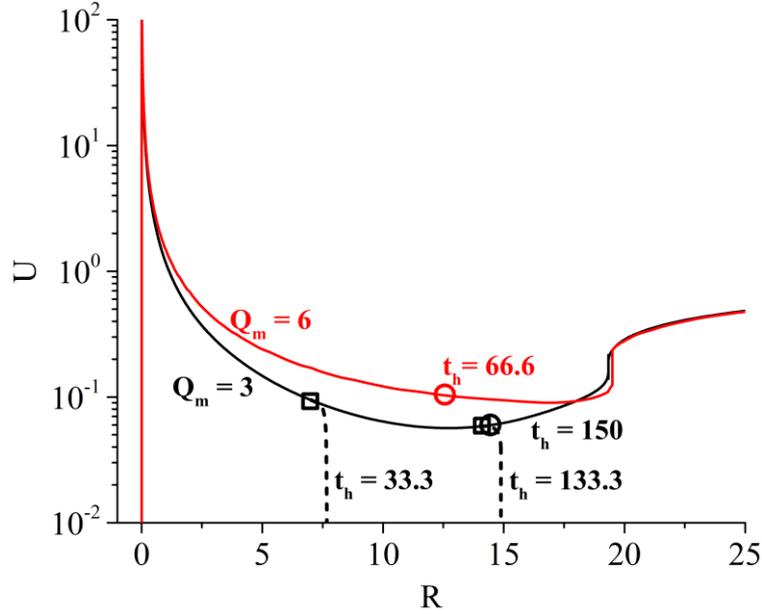

Figure 9. Change of flame propagation speed with flame radius for different ignition power and duration time. The Lewis number is $Le = 2$. The circles/squares represent the flame radius at the moment of external heating switching off, i.e., $t = t_h$, for successful/failing flame initiation.

First, we consider the same heating power of $Q_m = 3$, but different duration time of $t_h = 33.3$, 133.3 and 150. Figure 9 shows that the flame propagation speed abruptly reduces toward zero, implying flame extinction, when the external heating is switched off at $t_h = 33.3$ and 133.3. The increase in the heating duration time extends the radial location where flame extinction occurs. Though heat is still supplied from the kernel center towards the flame front for $t > t_h$, it gradually reduces as the flame propagates outwardly. When the heat generation from chemical reaction and heat conduction from the kernel center is overbalanced by the heat loss in the preheat zone, the flame structure cannot be maintained and extinction occurs (Chen & Ju 2008). For the same heating power of $Q_m = 3$, a slightly longer heating time, e.g., $t_h = 150$, leads to successful flame initiation. Therefore, the MIE is within the range of $400 < E_{min} < 450$ for $Q_m = 3$. However, when the heating power is doubled to $Q_m = 6$, successful ignition can be achieved with much shorter heating time of $t_h = 66.6$, implying that $E_{\min} < 400$ for $Q_m = 6$. This indicates that the MIE depends on the heating power, which will be discussed later.

In Fig 10(a), the temporal evolution of temperature profile is plotted for $Q_m = 3$ and $t_h = 33.3$. Interestingly, the flame can persistently propagate for a while before extinguishing occurs, indicated by the blue circle in Fig 9 with $R = 6.96$. Such phenomenon is identified as "memory effect" of heating (Joulin 1985, He 2000), which is attributed to the unsteady evolution of high temperature at the flame kernel within finite duration of time. Since the memory effect tends to drive the flame to propagate further outwardly, it is expected to affect the ignition and MIE. Figure 10(b) shows the results for $Q_m = 3$ and $t_h = 150$. When the central heating is switched of at $t_h = 150$, indicated by the black circle in Fig 9 with $R = 14.3$, the flame front can sustain expansion due to memory effect and arrive at the critical radius at $t = 211$ with a positive propagation speed. Passing the critical radius, the flame can propagate outwardly in a self-sustained manner and thereby successful ignition is achieved. Further increasing the heating power, i.e., $Q_m = 6$, the memory effect appears to be more sound because the distance between the radius where central heating is switched off, indicated by red circle, and the critical radius becomes even longer than that for $Q_m = 3$, as shown in Fig 9.

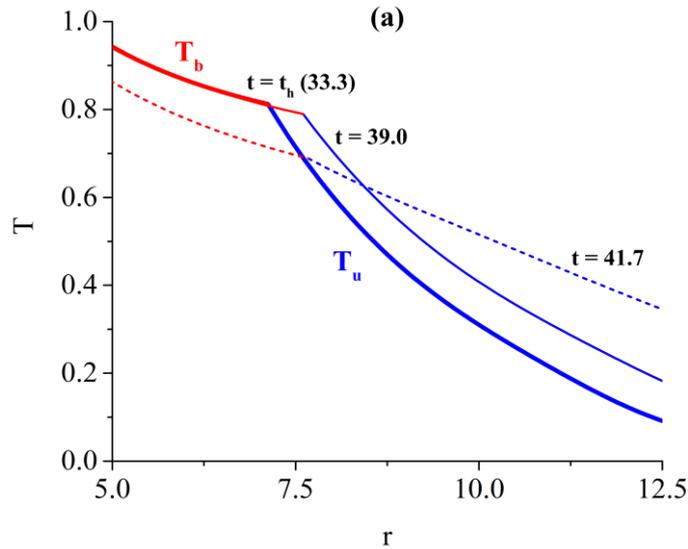

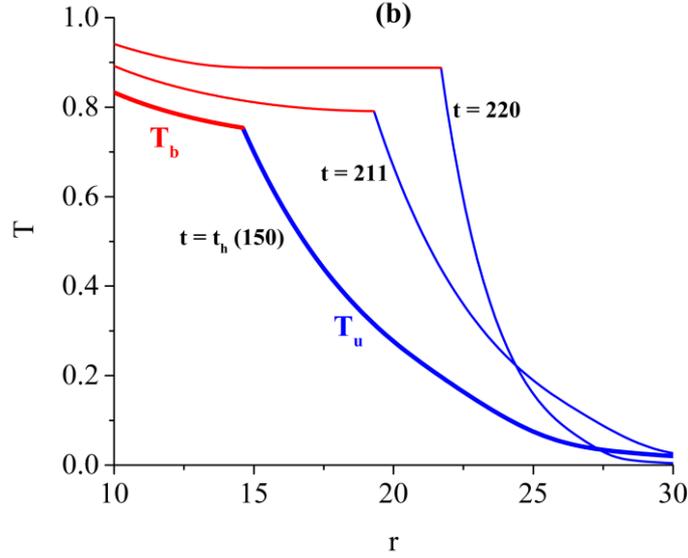

Figure 10. Temporal evolution of the temperature distributions for $Q_m = 3$ and $Le = 2$. The heating duration is (a) $t_h = 33.3$ and (b) $t_h = 150$. The thick lines correspond to $t = t_h$, while the thin lines denote the moments thereafter. The red and blue lines represent temperature in the burnt and unburned regions, respectively.

The change of MIE with heating power is presented in Fig 11. For comparison, the MIE predicted by the quasi-steady theory without considering memory effect, denoted as $E'_{min}$, is also plotted in Fig. 11. In the absence of characteristic time for temperature gradient evolution, the memory effect cannot be included in the quasi-steady theory. The dynamical behavior of flame front can be described by the U-R diagram, which regards the flame propagating speed as a function of flame front radius. Successful ignition requires the flame kernel to reach beyond the critical radius, i.e., $R > R_{cr}$. According to the definition of $U = dR/dt$, the moment for flame front arriving at the critical radius $R_{cr}$ can be evaluated by

$$t_{cr} = t_0 + \int_{R_0}^{R_{cr}} \frac{dR}{U} \tag{63}$$

where the integral on the right-hand side can be conducted with the knowledge of U-R diagram. We evaluate $t_{cr}$ based on the U-R relation determined through the transient formulation adopting constant heating power. The product of $t_{cr}$ and the heating power gives an overestimation to the MIE from the quasi-steady theory, i.e., $E'_{min} = t_{cr} Q_m$. The discrepancy between $E_{min}$ and $E'_{min}$ quantifies the impact of memory effect on the MIE.

Figure 11 shows that at relatively low heating power, the $E'_{min}$ agrees well with $E_{min}$, both of which rise abruptly as $Q_m$ approaching the critical value. The difference between $E_{min}$ and $E'_{min}$ becomes apparent as the heating power increases. For sufficiently intensive heating, $E_{min}$ considering memory effect tends to approach an asymptotic value, which grows cubically with the critical radius. However, $E'_{min}$ without considering memory effect changes with the heating power following an approximate scaling law, i.e., $E'_{min} \sim Q_m^{0.7}$ as indicated by the slope of the dashed lines in Fig 11. The growing discrepancy between the $E_{min}$ and $E'_{min}$ implies the increasing importance of memory effect in determining the MIE.

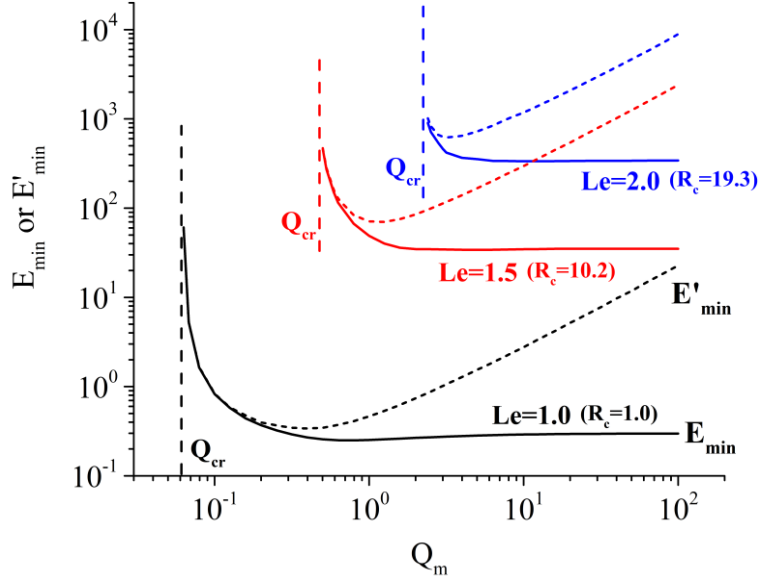

Figure 11. Change of the MIE with heating power for different Lewis numbers. The solid lines represent $E_{min}$ determined by the transient formulation with memory effect, while the dashed lines stand for $E'_{min}$ predicted by the quasi-steady theory without considering memory effect.

As mentioned before, the memory effect arises from the unsteady evolution of temperature gradient on the burnt side of the flame front, i.e., $(dT_b/dr)_{R^-}$ given by Eq. (47), after central heating switching off. The time change of $(dT_b/dr)_{R^-}$ is quantified by the $S$ function, given by Eq. (48), whose characteristic time can be obtained as

$$t_{bR} = \frac{R^2}{\hat{\mathcal{F}}_{bT}^2 \pi^2} \tag{64}$$

According to Eq. (36), the factor $\hat{\mathcal{F}}_{bT}$ changes with the flame front radius $R$ and propagation speed $U$, i.e.,

$$\hat{\mathcal{F}}_{bT} = \mathcal{F}_{bT}(\sigma_s = 1) = \frac{2}{\sqrt{\pi}} \frac{e^{-RU/2}\sqrt{RU/2}}{\text{erf}\left(\sqrt{RU/2}\right)} \tag{65}$$

We can quantify the memory effect by defining an extra distance of flame propagation driven by the memory effect,

$$R_{\text{me}} = t_{bR} U = \frac{R e^{RU}}{2\pi \left[\text{erf}\left(\sqrt{RU/2}\right)\right]^2} \tag{66}$$

Since the flame kernel establishing stage is extremely fast (see curve OA in Figs. 4), switching-off of external heating occurs during the stage of ignition-energy-supported flame kernel propagation. The presence of central heating overdrives the propagation of flame front, yielding that $U > 1$. Moreover, for mixtures with relatively large Lewis numbers, the critical radius for flame ignition tends to be considerably greater than thickness of planar flame thickness. Thereby, we hypothesize that the product of $RU$ at $t = t_h$, i.e., turning-off of central heating, could be regarded as a moderate-to-large quantity. According to Eq. (66) the extra distance $R_{\text{me}}$ appears as an increasing function of flame propagation sped.

With the increase in the heating power, the flame propagation is accelerated, and thereby the extra distance of flame propagation driven by memory effect, according to Eq. (66), becomes larger. Particularly, for sufficiently large heating power, the extra distance might be comparable with critical radius, i.e., $R_{\text{me}} \sim R_{cr}$, implying that the memory effect could play a dominant role in determining the MIE and thus leads to the exceedingly large discrepancy between $E_{\min}$ and $E'_{\min}$, as shown in Fig 11. Therefore, the MIE should be evaluated based on the transient formulation including the memory effect.

## 4. Concluding remarks

In this work, a fully transient formulation is proposed to analyze the development of a flame kernel in a quiescent mixture subject to external heating. The emphasis on the unsteady effects on ignition kernel propagation and MIE. Through a series of coordinate transformations, the conservation equations for energy and mass are converted into simple forms and solved analytically. Using the matching conditions at the flame front, we derive a pair of coupled implicit ordinary differential equations, whose solutions yield the time-dependent flame temperature, flame radius and flame propagation speed. Time scale analysis demonstrates that the present transient formulation is consistent with previous quasi-steady theory for stationary flame ball ($U = 0$) and for expanding flame approaching planar flame ($R \to \infty$). However, at intermediate radius with low to moderate propagating speed, i.e., $RU \sim O(1)$, the unsteady evolution time for temperature/mass fraction tends to comparable with that for flame propagation and thereby the unsteady effect could have discernible impacts upon the flame kernel development. The propagation speed for expanding flames at intermediate radius is found to be reduced by the unsteady effect.

Four stages involved in the flame initiation process subject to external heating are identified: the fast establishment of the ignition kernel, the ignition-energy-supported flame kernel propagation, unsteady transition of the flame kernel, and quasi-steady spherical flame propagation. The fundamental of each stage is clarified by examining the temporal and spatial variation of temperature/mass fraction distributions. The critical heating power predicted by quasi-steady theory appears to be linearly proportional to the cube of critical radius, i.e., $Q_{cr} \sim R_{cr}^3$. However, in transient formulation, the scaling law shall be revised to $Q_{cr} \sim R_{cr}^{3+\delta}$ with $\delta > 0$ due to unsteady evolution of temperature gradients across the flame front.

Furthermore, the present transient formulation can also deal with finite-duration central heating and thereby can predict the MIE. The MIE is found to be dependent on the heating power. For high heating power, the MIE predicted by the transient formulation approaches an asymptotic value while the MIE from the quasi-steady theory continuous increases. The memory effect of external heating induces the continuous propagation of flame front after the removal of heating source and thereby reduces the MIE. With the increase of heating power, the memory effect becomes stronger and thereby the discrepancy in the MIE predicted by the transient formulation and quasi-steady theory becomes larger.

It is noted that the present analysis is based on the assumption of one-step global chemistry and adiabatic flame propagation. In future studies, it would be interesting to consider simplified thermally sensitive intermediate kinetics (e.g., (Zhang & Chen 2011)) and radiative heat loss in the present transient formulation. Besides, here the flammable mixture is quiescent, and the flow caused by thermal expansion is not considered. It would be also interesting to take into account the uniform inlet flow and thermal expansion in future works.

## Acknowledgement

This work was supported by National Natural Science Foundation of China (nos. 51861135309 and 91741126).

# References

BUCKMASTER, J. & JOULIN, G. 1989 Radial propagation of premixed flames and t behavior. *Combustion and Flame* **78** (3-4), 275-286.

CHAMPION, M., DESHAIES, B. & JOULIN, G. 1988 Relative influences of convective and diffusive transports during spherical flame initiation. *Combustion and Flame* **74** (2), 161-170.

CHAMPION, M., DESHAIES, B., JOULIN, G. & KINOSHITA, K. 1986 Spherical flame initiation: Theory versus experiments for lean propane-air mixtures. *Combustion and Flame* **65** (3), 319-337.

CHEN, Z. 2017 Effects of radiation absorption on spherical flame propagation and radiation-induced uncertainty in laminar flame speed measurement. *Proceedings of the Combustion Institute* **36** (1), 1129-1136.

CHEN, Z., BURKE, M. P. & JU, Y. 2011 On the critical flame radius and minimum ignition energy for spherical flame initiation. *Proceedings of the Combustion Institute* **33** (1), 1219-1226.

CHEN, Z. & JU, Y. 2007 Theoretical analysis of the evolution from ignition kernel to flame ball and planar flame. *Combustion Theory and Modelling* **11** (3), 427-453.

CHEN, Z. & JU, Y. 2008 Combined effects of curvature, radiation, and stretch on the extinction of premixed tubular flames. *International Journal of Heat and Mass Transfer* **51** (25-26), 6118-6125.

CLAVIN, P. 2017 Quasi-isobaric ignition near the flammability limits. Flame balls and self-extinguishing flames. *Combustion and Flame* **175** 80-90.

CLAVIN, P. & SEARBY, G. (2016). *Combustion waves and fronts in flows: flames, shocks, detonations, ablation fronts and explosion of stars*, Cambridge University Press.

DESHAIES, B. & JOULIN, G. 1984 On the initiation of a spherical flame kernel. *Combustion Science and Technology* **37** (3-4), 99-116.

FERNÁNDEZ-TARRAZO, E., SÁNCHEZ-SANZ, M., SÁNCHEZ, A. L. & WILLIAMS, F. A. 2016 Minimum ignition energy of methanol–air mixtures. *Combustion and Flame* **171** 234-236.

HE, L. 2000 Critical conditions for spherical flame initiation in mixtures with high Lewis numbers. *Combustion Theory and Modelling* **4** (2), 159-172.

JACKSON, T., KAPILA, A. & STEWART, D. 1989 Evolution of a reaction center in an explosive material. *SIAM Journal on Applied Mathematics* **49** (2), 432-458.

JOULIN, G. 1985 Point-source initiation of lean spherical flames of light reactants: an asymptotic theory. *Combustion Science and Technology* **43** (1-2), 99-113.

KELLEY, A. P., JOMAAS, G. & LAW, C. K. 2009 Critical radius for sustained propagation of spark-ignited spherical flames. *Combustion and Flame* **156** (5), 1006-1013.

KURDYUMOV, V., BLASCO, J. & SÁNCHEZ PÉREZ, A. L. 2004 On the calculation of the minimum ignition energy. *Combustion and Flame* **136** (3), 394-397.

LAW, C. & SIRIGNANO, W. 1977 Unsteady droplet combustion with droplet heating—II: conduction limit. *Combustion and Flame* **28** 175-186.

LAW, C. K. (2006). *Combustion Physics*, Cambridge University Press.

RONNEY, P. D. (1989). On the mechanisms of flame propagation limits and extinguishment-processes at microgravity. *Symposium (International) on Combustion*.

RONNEY, P. D. 1990 Near-limit flame structures at low Lewis number. *Combustion and Flame* **82** (1), 1-14.

RONNEY, P. D. & SIVASHINSKY, G. I. 1989 A theoretical study of propagation and extinction of nonsteady spherical flame fronts. *SIAM Journal on Applied Mathematics* **49** (4), 1029-1046.

VÁZQUEZ-ESPI, C. & LIÑÁN, A. 2002 Thermal-diffusive ignition and flame initiation by a local energy source. *Combustion Theory Modelling* **6** 297-315.

VÁZQUEZ-ESPÍ, C. & LIÑÁN, A. 2001 Fast, non-diffusive ignition of a gaseous reacting mixture subject to a point energy source. *Combustion Theory and Modelling* **5** (3), 485-498.

VEERARAGAVAN, A. & CADOU, C. P. 2011 Flame speed predictions in planar micro/mesoscale combustors with conjugate heat transfer. *Combustion and Flame* **158** (11), 2178-2187.

WU, Y.-C. & CHEN, Z. 2012 Asymptotic analysis of outwardly propagating spherical flames. *Acta Mechanica Sinica* **28** (2), 359-366.

YA, Z., BARENBLATT, G., LIBROVICH, V. & MICHVILADZE, G. 1980 Mathematical theory of combustion and explosion, Moscow. *Science*.

YU, D. & CHEN, Z. 2020 Theoretical analysis on droplet vaporization at elevated temperatures and pressures. *International Journal of Heat and Mass Transfer* **164** 120542.

ZHANG, H. & CHEN, Z. 2011 Spherical flame initiation and propagation with thermally sensitive intermediate kinetics. *Combustion and Flame* **158** (8), 1520-1531.